\documentclass[a4paper]{article}

\usepackage{INTERSPEECH2021}
\usepackage{amsmath}
\usepackage{tabularx}
\usepackage{multirow}
\usepackage{makecell}

\title{Cross-Modal Knowledge Distillation Method for Automatic Cued Speech Recognition}
\name{Jianrong Wang$^1$, Ziyue Tang$^2$, Xuewei Li$^1$, Mei Yu$^1$, Qiang Fang$^3$, Li Liu$^{4*}$\thanks{* Corresponding author}}
\address{
  $^1$College of Intelligence and Computing, Tianjin University, Tianjin, China\\
  $^2$Tianjin International Engineering Institute, Tianjin University, Tianjin, China\\
  $^3$Institute of Linguistics, Chinese Academy of Social Science, Beijing, China\\
  $^4$Shenzhen Research Institute of Big Data, the Chinese University of Hong Kong, Shenzhen, China}
\email{liuli@cuhk.edu.cn}

\begin{document}

\maketitle
\begin{abstract}
Cued Speech (CS) is a visual communication system for the deaf or hearing impaired people. It combines lip movements with hand cues to obtain a complete phonetic repertoire. Current deep learning based methods on automatic CS recognition suffer from a common problem, which is the data scarcity. Until now, there are only two public single speaker datasets for French (238 sentences) and British English (97 sentences). In this work, we propose a cross-modal knowledge distillation method with teacher-student structure, which transfers audio speech information to CS to overcome the limited data problem. Firstly, we pretrain a teacher model for CS recognition with a large amount of open source audio speech data, and simultaneously pretrain the feature extractors for lips and hands using CS data. Then, we distill the knowledge from teacher model to the student model with frame-level and sequence-level distillation strategies. Importantly, for frame-level, we exploit multi-task learning to weigh losses automatically, to obtain the balance coefficient. Besides, we establish a five-speaker British English CS dataset for the first time. The proposed method is evaluated on French and British English CS datasets, showing superior CS recognition performance to the state-of-the-art (SOTA) by a large margin.
\end{abstract}
\noindent\textbf{Index Terms}: Cued Speech, Cross-modal knowledge distillation, Teacher-student structure, Cued Speech recognition

\section{Introduction}

\begin{figure}[t]
  \centering
  \setlength{\belowcaptionskip}{-0.1cm}
  \includegraphics[width=\linewidth]{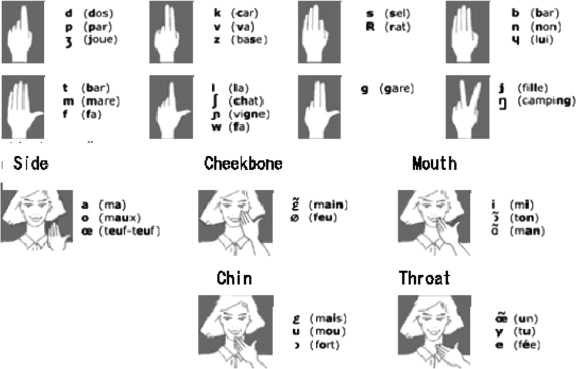}
  \caption{Chart of the French Cued Speech (from \cite{heracleous2010cued}).}
  \label{fig:fr-cs}
\end{figure}

Cued Speech (CS) is a visual mode of communication for hearing impaired people proposed by Cornett \cite{cornett1967cued} in 1967. It is to solve the confusion of lip reading. CS uses the shape and position of hand to assist in lip reading to make visual communication easier and more efficient for hearing impaired people. The hand shape is used to encode consonants, and the hand position is used to encode vowels (see Figure~\ref{fig:fr-cs}).

Currently, there are some automatic CS recognition works \cite{heracleous2010cued, heracleous2012continuous, liu2018visual, papadimitriou2021fully, liu2020re} based on two small scale single-speaker datasets (one is in French \cite{liu2018automatic} and the other is in British English \cite{liu2019automatic, liu2019novel}). The tandem architecture in \cite{liu2018visual} achieved 62\% accuracy on the French CS dataset with single speaker and 238 French sentences. More recently, \cite{papadimitriou2021fully} proposed a deep sequence learning approach, which consists of an image learner based on convolutional neural networks (CNNs) \cite{lecun1995convolutional} and a fully convolutional encoder-decoder. It was evaluated on the British English CS dataset for the first time achieving 63.75\% with single speaker and 97 British English sentences. These two works were both focus on continues CS recognition with no artificial marks \cite{heracleous2010cued,heracleous2012continuous,aboutabit2007reconnaissance,burger2005cued} but their performance were still limited by the data scarcity, which leads to the overfitting in a deep neural network training.
\begin{figure*}[t]
  \centering
  \setlength{\belowcaptionskip}{-0.1cm}
  \includegraphics[width=5 in]{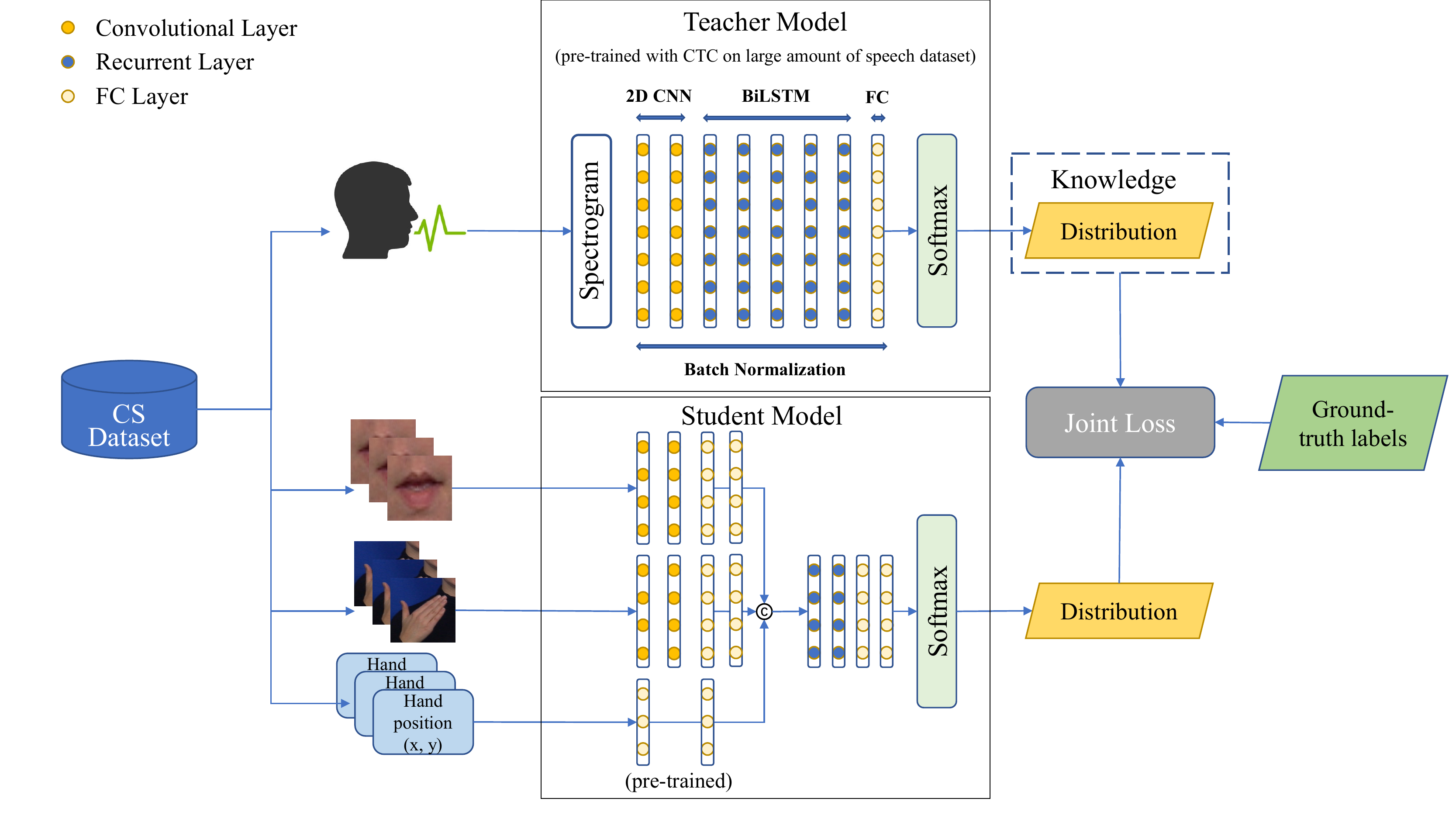}
  \caption{Teacher-student structure for automatic CS recognition.}
  \label{fig:kd-model}
\end{figure*}

Knowledge distillation with the teacher-student structure can transfer and preserve the cross-modality knowledge successfully \cite{thoker2019cross, luo2018graph, crowley2018moonshine, hao2019spatiotemporal, wu2019multi, zhang2020knowledge}. It has been successfully applied in many fields. For example, in \cite{gupta2016cross}, Gupta et al. transferred the knowledge from the teacher model to the student model relying on unlabeled paired samples involving both RGB and depth images. Zhao et al. \cite{zhao2018through} used synchronized radio signals and camera images to transfer the knowledge across modalities for radio-based human pose estimation. Thoker and Gall \cite{thoker2019cross} transferred the knowledge obtained from RGB videos to a skeleton-based human action recognition model.

In this study, considering the limited size and multi-modality (including audios and videos) of CS data, we innovatively propose a cross-modal knowledge distillation architecture, which  transfers the knowledge learned from audio speech to visual-based CS (see Figure~\ref{fig:kd-model}). Although the size of CS dataset is limited, many acoustic speech datasets are publicly available. What's more, CS and audio speech are both phoneme-level coding thus they have the same phoneme semantics. It motivates us to exploit cross-modal distillation in CS recognition. Firstly, we pretrain a teacher model for phoneme recognition with a large amount of open source audio speech data, and simultaneously pretrain feature extractors for lips, hand shape and hand position. Then, we distill the speech knowledge from the large teacher model into the small student model to estimate the parameters of bi-directional long short-term memory (BiLSTM) \cite{graves2013speech} in the student network with frame-level and sequence-level \cite{huang2018knowledge, takashima2019investigation} distillation strategies. For frame-level distillation, the weights of losses are learned automatically based on multi-task learning, and then they are used to compute the balance coefficient (\textit{i.e.}, hyperparameters used to balance the gradient among loss functions).

Compared with \cite{liu2018visual}, we achieve significant CS phoneme recognition accuracy improvement of 12.2\% in frame-level (17\% in sequence-level) for French task. Compared with the state-of-the-art (SOTA) \cite{papadimitriou2021fully}, we improve 3.32\% in frame-level (8.12\% in sequence-level) for French CS and 1.15\% in frame-level (7.95\% in sequence-level) for British English CS with single speaker.

In summary, the key contributions of this work are:
\begin{enumerate}
\item This is the first attempt to apply cross-modal knowledge distillation method to automatic CS recognition to overcome the problem of overfitting caused by limited size of CS dataset.
\item To obtain a better model automatically, we exploit multi-task learning to weigh losses, and then compute the balance coefficient to balance the gradient among loss functions.
\item We establish the first five-speaker British English CS dataset with 390 English consecutive sentences.
\end{enumerate}

\section{Methodology}

In this section, we introduce our method, which includes teacher network and student network in the cross-modal knowledge distillation architecture, as well as frame-level and sequence-level distillation strategies.

\subsection{DeepSpeech2 teacher network}

DeepSpeech2 \cite{amodei2016deep} is an end-to-end deep learning approach for automatic speech recognition (ASR) with good generalization performance. The architecture of DeepSpeech2 we used is shown as the Teacher Model in Figure~\ref{fig:kd-model}. It consists of two 2D convolutional layers followed by five BiLSTM layers with batch normalization \cite{ioffe2015batch} and a fully connected layer (FC). It is trained with the connectionist temporal classification (CTC) \cite{graves2006connectionist} loss function to predict audio speech transcriptions.

To use DeepSpeech2 as teacher model for CS recognition, we translate sentence label into CS phoneme sequence. Let $x^{(i)}$ is a time-series, where each time-slice is a vector of audio features, and $y^{(i)}$ is the corresponding label. The inputs of the network are spectrogram of power normalized audio clips, and the outputs are the phonemes. At each output time-step $t$, BiLSTM makes a prediction over phonemes, $p(l_t|x)$ where $l_t \in \{blank,\Gamma\}$. $\Gamma$ is the set of CS phonemes that are different in French and British English. We represent the CTC loss function as $L(x,y;\theta)$, where $(x,y)$ is an input-output pair, and $\theta$ is the current parameters of the network. The network parameters are updated making use of the derivative $\triangledown_\theta L(x,y;\theta)$.

\subsection{CNN-BiLSTM student network}

As shown in Figure~\ref{fig:kd-model}, the architecture of the student model consists of two pre-trained CNNs and one pre-trained artificial neural network (ANN) \cite{schalkoff1997artificial}, followed by two BiLSTM layers and two FC layers. CNNs act as feature extractors. Each focuses on lips or hands separately, trained with a set of 8 visemes \cite{benoit1992set} as targets for lips and 8 shapes for hands. The details of the CNNs and ANN network structure are the same as \cite{liu2018visual}. After extracting features based on CNNs and ANN, the three stream features are concatenated in a single feature vector as the input of BiLSTM.

To verify the effectiveness of knowledge distillation for CS recognition in frame-level and sequence-level, we evaluate the student model trained by minimizing Cross Entropy (CE) loss and CTC loss, respectively.

\subsection{Knowledge distillation strategy}

We consider two distillation strategies \textit{i.e.}, frame-level and sequence-level. The standard knowledge distillation applied to frame-level prediction should be effective for CS recognition, and a novel sequence-level knowledge distillation strategy might further improve performance.

\subsubsection{Frame-level distillation strategy}
\label{frame-level section}

For frame-level distillation strategy, we try an existing general method firstly. Considering a generalized $softmax$ function \cite{hinton2015distilling}:
\begin{equation}
  q_i = \frac{exp(z_i/T)}{\sum_j exp(z_j/T)},
  \label{eq1}
\end{equation}
where $z_i$ is the logit from network output layer, $q_i$ is the class probability corresponding to each $z_i$, and $T$ is the temperature coefficient that is normally set to 1. Both the soft targets from the teacher and the ground-truth label are of great importance for improving the performance of the student model. Therefore, we consider Kullback-Leibler Divergence (KL) as distillation loss and CE as student loss as follow:
\begin{equation}
  L(x_t,x_s;W) = \alpha KL(P_{(T)},Q_{(T)})*T^2 + (1-\alpha)CE(Y,Q_{(1)}),
  \label{eq:JLF1}
\end{equation}
where $x_t$ is the CS audio input of the pre-trained teacher model and $x_s$ is the input of the student model which is concatenated feature vector. $W$ are the parameters of the student model, and $\alpha$ is a regulated parameter. $P_{(T)}$ and $Q_{(T)}$ are the $softmax$ output of teacher model and student model, respectively when temperature is $T$. $Y$ is the one-hot ground-truth label, and $Q_{(1)}$ is the $softmax$ output of student model when the temperature is 1. For this loss function, we tune $T$ and $\alpha$ manually to find a model with higher CS phoneme recognition accuracy.

\begin{figure}[t]
  \centering
  \includegraphics[width=0.7\linewidth]{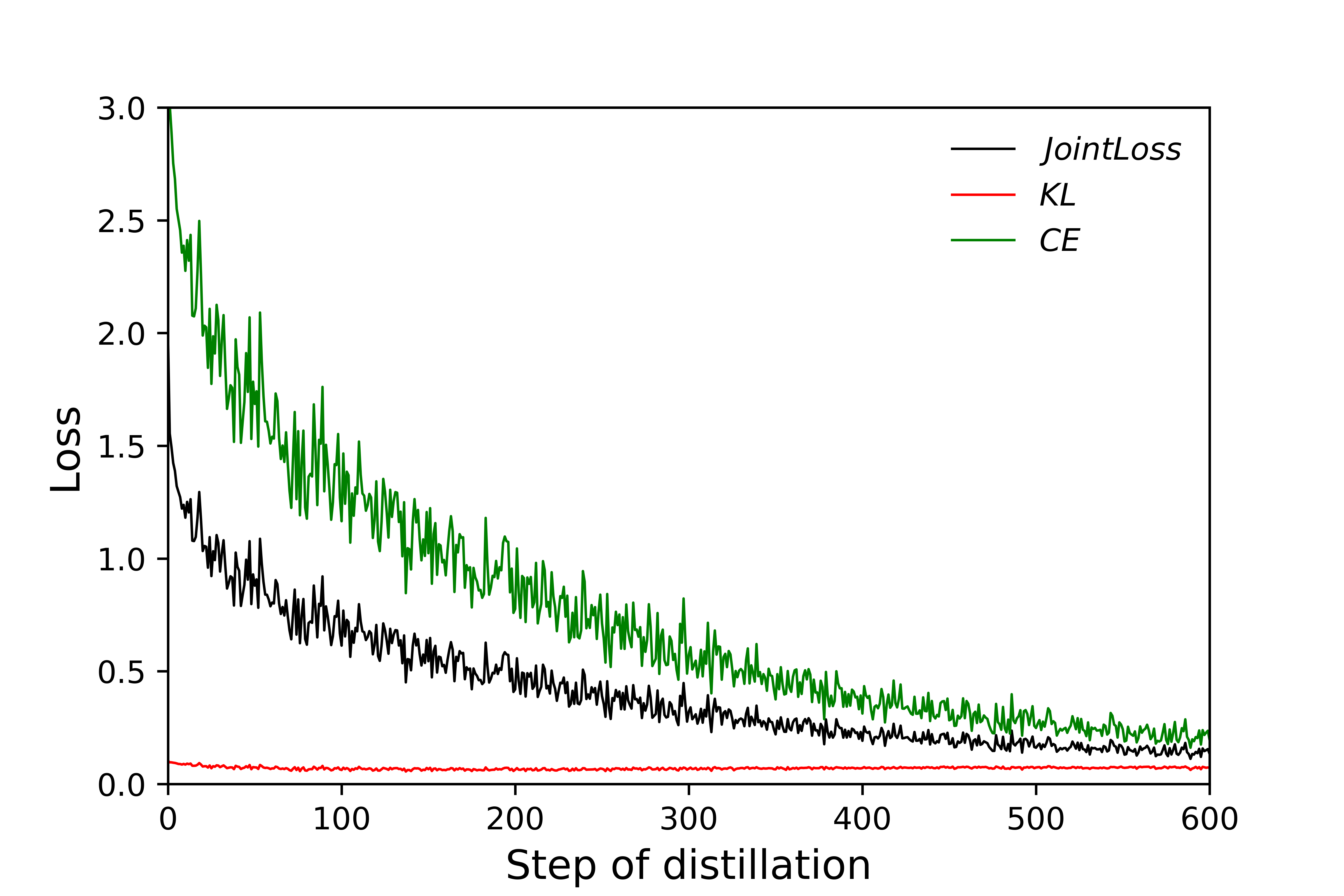}
  \caption{Comparison of gradient descent for KL, CE and their joint loss when $T=1$ and $\alpha = 0.5$ in Equation (\ref{eq:JLF1}).}
  \label{fig:loss_KL_CE_only}
\end{figure}

The loss reduction curve is shown in the Figure~\ref{fig:loss_KL_CE_only} when $T$ is set to 1 and $\alpha$ is set to 0.5. We observe that the gradient of KL is almost 0, which means that the knowledge of the teacher model is not effectively distilled to the student model, and there is an imbalance in the gradient between KL and CE. To avoid tuning $T$ and $\alpha$ manually, we derive a multi-task joint loss function as Equation (\ref{eq:mtl}) based on the task uncertainty.
\begin{equation}
  L_{mtl} = \frac{1}{\sigma_1 ^2}KL(P,Q) + \frac{1}{\sigma_2 ^2}CE(Y,Q) + log\sigma_1 + log\sigma_2,
  \label{eq:mtl}
\end{equation}
where $T$ and $\alpha$ are not considered anymore, and we automatically learn the observation noise scalar $\sigma_1$ and $\sigma_2$. As reported in \cite{kendall2018multi}, the task uncertainty captures the relative confidence between tasks, reflecting the uncertainty inherent to the regression or classification task. Therefore, in our two classification tasks (\textit{i.e.}, the student model learns from the teacher model and learns from the ground-truth) of the distillation process, uncertainty is used as a basis to weigh losses.

Further, to balance the loss gradient between these two tasks, the loss joint function based on balance coefficient (\textit{i.e.}, $a$) is proposed as:
\begin{equation}
  L(x_t,x_s;W) =  \frac{1}{2}(a\times KL(P_{(1)},Q_{(1)}) + CE(Y,Q_{(1)})),
  \label{eq:JLF3}
\end{equation}
We obtain the balance coefficient through $a=\frac{\sigma_2 ^2}{\sigma_1 ^2}$.

\subsubsection{Sequence-level distillation strategy}

Similarly, for sequential distillation, both the soft targets from the teacher and the ground-truth label are of great importance for improving the performance of the student model. Therefore, for student network, a distillation loss that assists CTC is explored. Rather than using KL divergence, we apply cosine similarity (cos) as the distillation loss, since it pays more attention to the difference in the direction between the two vectors. It is shown as follows:
\begin{equation}
  L(x_t,x_s;W) = \frac{1}{2} (1 - cos(S_s,S_t) + CTC(Y,logQ)),
  \label{eq:sequence JLF}
\end{equation}
where $x_t$, $x_s$, $W$ and $Q$ are the same as described in section \ref{frame-level section}. $S_s$ and $S_t$ are the frame-based logits before the $softmax$ layer of the student and teacher model, respectively. $Y$ is corresponding transcript.

\section{Experiments}

\subsection{Datasets}
\textbf{\textit{CS datasets for student model}}
 We use a public French CS dataset \cite{liu2018automatic} and the newly built British English CS dataset with 5 speakers specially for this work. The detail of the public single-speaker French can refer to \cite{liu2018automatic}. Here, we mainly introduce the new multi-speaker British English CS dataset.

A single-speaker (named CA) British English CS dataset \cite{liu2019automatic} which contains 97 sentences. Now, the dataset has been expanded with data for other four CS interpreters named EM, KA, LD and VK, respectively. There are 390 British English audios and videos of all five speakers in total. The RGB video image recorded at 25fps. In the British English CS system, the 12 monophthongs are encoded by four hand positions and the 8 diphthongs are encoded by four hand slips, while the 24 consonants are encoded by eight hand shapes.

\noindent \textbf{\textit{Datasets for teacher model pre-training}}
 We train two teacher models for distillation with the French speech dataset and English speech dataset in Common Voice \cite{ardila2019common}.

Common Voice is an open source project launched by Mozilla including French and English speech data. The total number of verification hours of French version is 350h and English version is 1118h. We convert the French sentence text into French CS phonemes using Lliaphon  \cite{bechet2001lia}. Then British English phonetic transcript is obtained by eSpeak (\textit{i.e.}, a compact open source software speech synthesizer for English and other languages). It should be mentioned that the English datasets only contains 8\% of British English, and the remaining 92\% consists of American English and dialects. 

\begin{table*}[th]
  \caption{Performance comparison in Acc(\%) in frame-level distillation. Fully Conv \cite{papadimitriou2021fully} is the SOTA. $\rm JLF1$ is the best performance obtained by minimizing the joint loss function represented by Equation (\ref{eq:JLF1}). $\rm JLF2_{mtl}$ represents the performance of the optimal model obtained by Equation (\ref{eq:mtl}).
   $\rm JLF3$ is the performance obtained by Equation (\ref{eq:JLF3}).}
  \label{tab:frame-level results}
  \centering
  \begin{tabularx}{\textwidth}{l|X|X|X}
    \toprule
    Dataset & French & \makecell{Single-speaker\\(British)} & \makecell{Multi-speaker\\(British)}                \\
 	\midrule
	CNN-HMM \cite{liu2018visual} & 62\% & --- & --- \\
	\midrule
	Fully Conv \cite{papadimitriou2021fully} & 70.88\% & 63.75\% & --- \\
	\midrule
	Student CE & 64.4\% & 52.5\% & 62.3\%                 \\
	\midrule
	$\rm JLF1$ & 72.5\% ($T=5,\alpha=0.5$) & 61.3\% ($T=5,\alpha=0.7$) & 65.7\% ($T=5,\alpha=0.7$) \\
	\midrule
	$\rm JLF2_{mtl}$ & 72.5\% ($\sigma_1=0.37,\sigma_2=1.20$) & 63.1\% ($\sigma_1=0.59,		\sigma_2=1.02$) &  68.5\% ($\sigma_1=0.23,\sigma_2=0.74$) \\
	\midrule
	$\rm JLF3$ & \textbf{74.2\%} ($a=\frac{\sigma_2 ^2}{\sigma_1 ^2} \approx 10$) & \textbf{64.9\%} ($a=\frac{\sigma_2 ^2}{\sigma_1 ^2} \approx 3$) & \textbf{69.7\%} ($a=\frac{\sigma_2 ^2}{\sigma_1 ^2} \approx 10$) \\
	\bottomrule
  \end{tabularx}
\end{table*}

\subsection{Protocol and metrics}

In our experiments, each CS dataset for student model uses 80\% of the data for training, 10\% of the data for validation, and the other 10\% for testing. These three parts of data do not overlap. All models are evaluated in CS phoneme accuracy (Acc) which is defined as $Acc (\%) = 1-PER (\%)$.  PER is the phoneme error rate.

\subsection{Result and analysis}

We evaluate the teacher model, as well as the student model trained with distillation in frame-level and sequence-level on CS datasets.

\subsubsection{Evaluation on teacher model}

The phoneme recognition accuracy of the teacher model is 83.7\% on French CS and 70.1\% on the new British English CS audio speech data. We hypothesize that the low performance of the English teacher model is caused by the limited British English data (only 8\% in Common Voice French dataset).

\subsubsection{Evaluation on student model}

\textbf{\textit{Frame-level distillation}}
 For frame-level knowledge distillation, we evaluate three loss functions (\textit{i.e.}, Equation (\ref{eq:JLF1}), (\ref{eq:mtl}) and (\ref{eq:JLF3})). In Table~\ref{tab:frame-level results}, we find out that the knowledge distillation method bring considerable improvement compared with training student model with baseline CE. On each CS dataset, $\rm JLF3$ is the best result. We achieve Acc improvement of 12.2\% compared with CNN-HMM \cite{liu2018visual},  and 3.32\% compared with SOTA \cite{papadimitriou2021fully} in French CS task. In single-speaker British CS task, we improve 1.15\% Acc compared with SOTA. For the new dataset with multi speakers, $\rm JLF3$ achieves Acc of 69.7\%.

\begin{figure}[t]
  \centering
  \setlength{\belowcaptionskip}{-0.1cm}
  \includegraphics[width=0.9\linewidth]{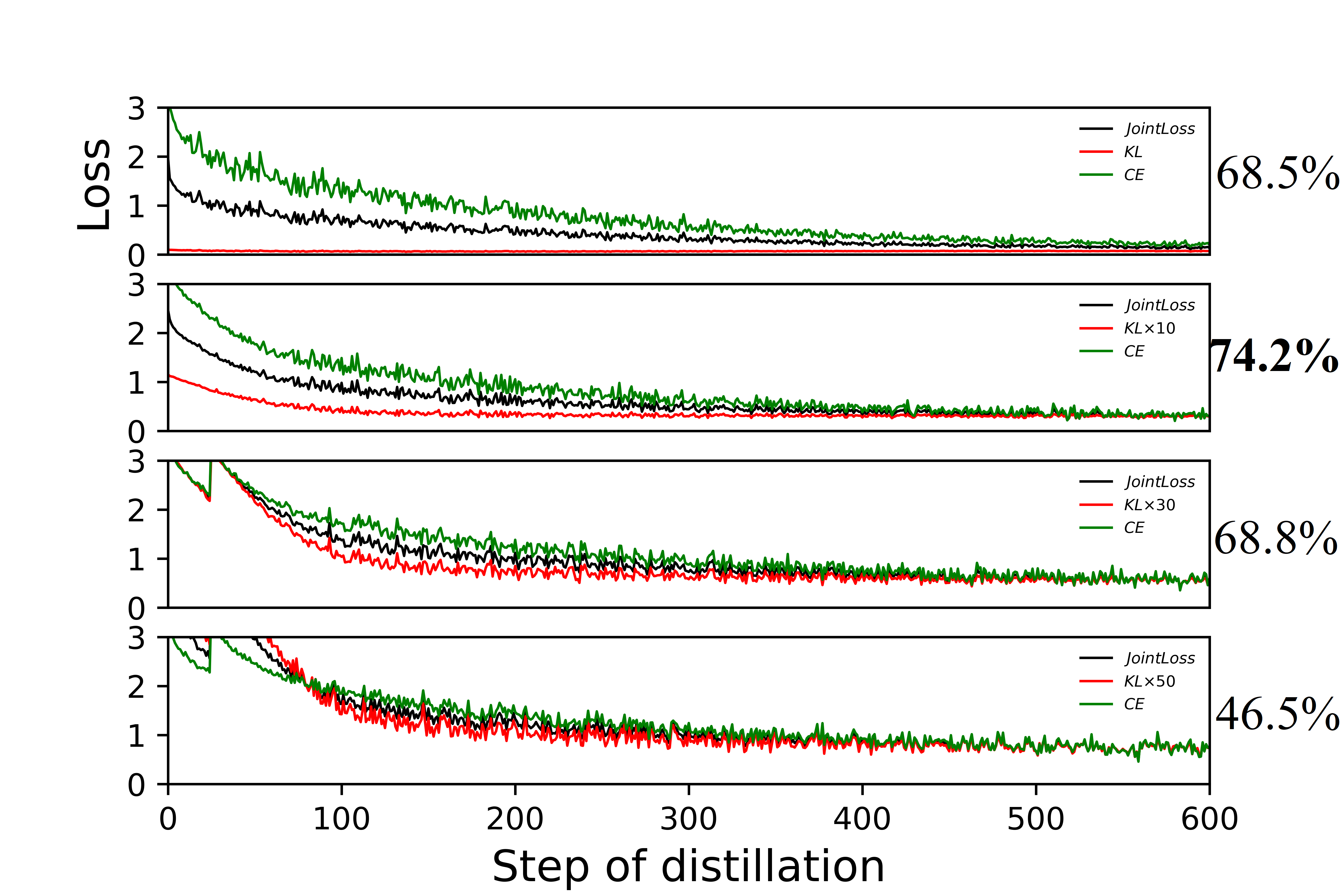}
  \caption{Comparison of gradient descent among KL, CE and their joint loss with different balance coefficients in French CS recognition task.}
  \label{fig:loss_curves_4}
\end{figure}

More balance coefficients are explored to show the effectiveness of the joint loss function we proposed (see Equation (\ref{eq:JLF3})). From Figure~\ref{fig:loss_curves_4}, it can be observed that the closer the gradient between the KL and CE at the beginning of the training step, the more efficient the distillation.

\noindent \textbf{\textit{Sequence-level distillation}}
 As demonstrated in Table~\ref{tab:sequence-level results}, in addition to multi-speaker, as an auxiliary loss, cos performs better than training the student model with CTC alone. The English teacher model (70.1\%) is not better than the student model trained by CTC (78.6\%), so that it is not enough to play a positive role in guiding the student model.
\begin{table}[th]
  \caption{Performance comparison by Acc(\%) in sequence-level distillation. Fully Conv \cite{papadimitriou2021fully} is the SOTA.}
  \label{tab:sequence-level results}
  \centering
  \begin{tabular}{l|l|l|l}
    \toprule
    Dataset & French & \makecell{Single-speaker\\(British)} & \makecell{Multi-speaker\\(British)}                \\
 	\midrule
	CNN-HMM & 62\% & --- & --- \\
	\midrule
	Fully Conv & 70.88\% & 63.75\% & --- \\
	\midrule
	Student CTC & 77\% & 68.6\% & 78.6\%                 \\
	\midrule
	KL+CTC & 75\% & 68.4\% & 75.2\% \\
	\midrule
	cos+CTC & \textbf{79\%} & \textbf{71.4\%} & 77.5\% \\
	\bottomrule
  \end{tabular}
\end{table}

To conclude, we verify the effectiveness of knowledge distillation, the joint loss function based multi-task learning and the balance coefficient in CS recognition with limited data size. Besides, the performance on the multi-speaker CS dataset shows good generalization performance for different speakers.

\section{Conclusions}

In this work, we proposed a cross-modal knowledge distillation method including two structures, a teacher model and a CNN-BiLSTM student model with two distillation strategies for CS recognition on limited size of datasets. We highlighted the effectiveness of our method and the importance of weighing losses based multi-task learning as well as the balance coefficient for the frame-level distillation gradient descent. The comparative evaluation demonstrated that the proposed method achieves new SOTA on the CS recognition in the multi-speaker scenario. As for future work, we will be engaged in improving the teacher and student model by language models to decrease the insertion errors.

\section{Acknowledgements}
This work was supported by the GuangDong Basic and Applied Basic Research Foundation (No. 2020A1515110376) and the National Natural Science Foundation of China (grant No. 61977049). The authors would like to thank the professional CS speakers from Cued Speech UK for the British English CS dataset recording.
\bibliographystyle{IEEEtran}

\bibliography{mybib}

\end{document}